\documentclass[vecphys]{svmult}

\usepackage{makeidx}         
\usepackage{graphicx}        
\usepackage{multicol}        
\usepackage[bottom]{footmisc}

\makeindex             

\begin{document}

\title*{Dust Processing and Mineralogy in Protoplanetary Accretion Disks}
\titlerunning{Chapter III: Dust in Protoplanetary Disks}
\author{Thomas Henning \and Gwendolyn Meeus}
\institute{ Max Planck Institute for Astronomy, K\"onigstuhl 17,
D-69117 Heidelberg, Germany \texttt{henning@mpia.de} \and
Universidad Aut\'onoma de Madrid, Departamento de F\'isica Te\'orica
C-XV, 28049 Madrid, Spain \texttt{gwendolyn.meeus@uam.es}
}
%
%
\maketitle


In this chapter we discuss the different dust components a
protoplanetary disk is made of with a special emphasis on grain
composition, size and structure. The paper will highlight the role
these dust grains play in protoplanetary disks surrounding young
stars, as well as observational results supporting this knowledge.
First, the path dust travels from the interstellar medium into the
circumstellar disk is described. Then dust condensation sequences
from the gas are introduced, to determine the most likely species
that occur in a disk. The characteristics of silicates are handled
in detail: composition, lattice structure, magnesium to iron ratio
and spectral features. The other main dust-forming component of
the interstellar medium, carbon, is presented in its many forms,
from molecules to more complex grains. Observational evidence for
polycyclic aromatic hydrocarbons (PAHs) is given for both young
stars and solar system material. We show how light scattering
theory and laboratory data can be used to provide the optical
properties of dust grains. From the observer's point of view, we
discuss how infrared spectra can be used to derive dust
properties, and present the main spectral analysis methods
currently used and their limitations. Observational results,
determining the dust properties in protoplanetary disks, are
given: first for the bright intermediate-mass Herbig Ae/Be stars,
and then for the lower-mass Tauri stars and brown dwarfs. Here we
present results from the space observatories {\it ISO} and {\it
Spitzer}, as well as from the mid-infrared interferometer VLTI,
and summarise the main findings. We discuss observational evidence
for grain growth in both Herbig Ae/Be and T Tauri stars, and its
relation with spectral type and dust settling.  We conclude with
an outlook on future space missions that will open new windows,
towards longer wavelengths and even fainter objects.

\section{Introduction}

Dust grains dominate the opacity in protoplanetary disks whenever
they are present. This implies that their radiation properties
play a crucial role in determining the temperature and density
structure of these disks. The initial population of
(sub)micron-sized particles evolves over time towards
planetesimals, eventually providing the building blocks for
terrestrial planets.

The dust grains shield the interior of protoplanetary disks from
energetic cosmic particles and stellar X-ray radiation and provide
the surface for electron recombination. This regulates the
ionization structure of disks, which is an important ingredient
for the magneto-rotational instability to operate and to drive
angular momentum transport. The dust particles are equally
important for disk chemistry because chemistry on grain surfaces
leads to the formation of molecular ices and, possibly, complex
organic molecules, which will enter the gas phase when their
evaporation temperature is reached.

Infrared spectroscopy is a powerful tool to characterize the
properties of protoplanetary dust. With ground-based telescopes,
the {\it Infrared Space Observatory ISO} and the {\it Spitzer
Space Telescope}, an enormous amount of data has been obtained to
characterize the mineralogy of disks around a variety of objects,
ranging in luminosity from Herbig Ae/Be stars to T Tauri stars,
and even brown dwarfs. These data allow us to put constraints on
the chemical composition and amorphous/crystalline state of the
dust particles, and to address questions such as radial
distribution and mixing processes. Mid-infrared long-baseline
interferometry is starting to contribute as well to our
understanding of the structural properties of dust in the
different radial zones of protoplanetary disks.


\section{Dust Components in Protoplanetary Disks}
\label{s_dust}
\subsection{General Overview}

As the infalling disk material originates from the interstellar
medium (ISM), more precisely from the parental molecular cloud
core, the initial dust composition in a protoplanetary accretion
disk is assumed to be similar to the molecular cloud dust
composition. It can be slightly altered from this composition, as
volatile molecular ices could evaporate during the passage of the
accretion shock front, oxygen could convert into water and quartz
(SiO$_2$) could form from silicon atoms. For a detailed discussion
of the various dust populations in space, including dust in
molecular clouds, we refer to the review by Dorschner \& Henning
(1995). The most abundant cosmic dust species are compounds of O,
Si, Mg, Fe, and C: silicates and carbonaceous dust.

Depending on the angular momentum of the material, molecular cloud
dust will accrete onto the disk at different radial distances from
the star, which may influence subsequent grain evolution
(Dullemond et al. 2006). In protoplanetary disks, a whole range of
modification processes are expected to occur, ranging from thermal
annealing in hot regions of the inner disk, ion irradiation by
stellar flares, X-ray and UV irradiation, destruction of
carbonaceous dust by oxidation close to the central star,
equilibration with the gas through sublimation-condensation
processes and solid-phase reactions, as well as molecular ice
formation in the outer disk. In addition, grain growth and both
radial and vertical mixing processes need to be considered,
leading to the expectation that the grain composition changes both
over time and with radial distance from the star, so that a
relatively large diversity of dust compositions can be expected in
protoplanetary disks.

As a reference for the composition of protoplanetary dust in the
outer disk regions, the dust model introduced by Pollack et al.
(1994) is widely used. It is based on the solar elemental
composition and the ISM gas depletion pattern, the composition of
primitive solar system material, including the mass-spectroscopy
results of the space probes to comet Halley, and theoretical
considerations. The model contains the iron-magnesium silicate
minerals olivines and pyroxenes, quartz, metallic iron, troilite
(FeS), volatile and refractory organics, as well as water ice in
the outer disk. The authors already noted that the silicates are
certainly mixtures of amorphous and crystalline silicates, with
the amorphous silicates dominating. In fact, they used the optical
constants of amorphous silicates at mid-infrared wavelengths as a
pragmatic choice. The dust model divides the products of the most
abundant dust-forming elements O, C, Si, Mg, Fe, S, and N into
three categories: gases, molecular ices, and refractory grains.
Based on the dust composition and the optical properties of these
various components, Pollack et al. (1994) also derived dust
opacities for disks. They were further improved by including dust
aggregates (Henning \& Stognienko 1996), as well as updated
optical data for the various materials (Semenov et al. 2003).

Molecular ices only exist in the cold outer disk, while FeS forms
from Fe and H$_2$S at condensation temperatures of 680 K. Besides
water ice, also CO, CO$_2$, NH$_3$, CH$_4$, and CH$_3$OH ices have
been detected in the infrared spectra of disks (Pontoppidan et al.
2005, Zasowski et al. 2008).

The exact fractional abundance of the various solids remains an
open question. In general, we would expect a radial variation of
the dust composition, as already mentioned for Fe/FeS and
molecular ices. As another example, carbonaceous dust should be
destroyed in the inner disk when evolving towards an
equilibrium dust mixture under oxygen-rich conditions.

The bulk of the ISM/molecular cloud dust consists of amorphous
silicates and amorphous carbonaceous material.  ISO observations
around 10 micron have shown that most of the silicates ($\geq$
98\% by mass) in the ISM have an amorphous structure (e.g. Kemper
et al. 2004). In contrast, infrared spectroscopy of Herbig Ae/Be
stars, T Tauri stars and brown dwarfs has demonstrated that a
significant fraction of the dust in protoplanetary disks is in a
crystalline state, implying that these crystals have been formed
inside the disks (see Sect.~\ref{s_minera}). Amorphous silicates
only crystallize at relatively high temperatures through thermal
annealing processes (e.g. Fabian et al. 2000), suggesting that
these materials have experienced high temperatures in the disks
(typically 800--1000~K).

In an actively accreting disk, the main accretion flow points to
the star and most of the dust will be destroyed by sublimation and
subsequently incorporated in the star. Once the main
accretion phase has terminated and the star has grown close to its
final mass, the material will only slowly move inwards. In both cases,
the dust grains will experience an increase in temperature when
approaching the star. This will lead to both (1) annealing of the
amorphous material into a more crystalline structure and (2)
chemical processing through evaporation and re-condensation, herewith
changing the abundance of the different species. It is important
to remember that the dust species are the main source of the
opacities, which determine the disk structure. So when chemical
processing or sublimation causes a certain dust species to
disappear, a change in the opacities will occur, with consequences
for the structure of the disk.

\subsection{Condensation of dust}

Condensation sequences of dust from the gas phase are often based
on chemical equilibrium calculations, starting from the early
studies by Larimer (1967), Grossman (1972), Lattimer et al. (1978)
to the more recent investigations by Gail (1998) and Krot (2000).
Although the gas-dust mixture may often not be in a state of
chemical equilibrium, these calculations are a useful tool to
predict which dust species can be expected in the considered
elemental gas mixture.

Gail (1998; see also Gail 2003 for a comprehensive discussion)
proposed a scheme to determine the stable dust materials expected
to be present in a disk. It starts from the - vaporized into the
gas phase - dust mixture from Pollack et al. (1994), and is based
on chemical equilibrium considerations. The most important
parameters in this scheme are the temperature and the pressure,
which critically depend on the location in the disk; as a
consequence, the stability of a certain dust species (hence its
presence) will be radially dependent. To summarize Gail's results,
we will give an overview of the condensates that are expected to
be stable at a certain temperature, starting from the outer disk
region and moving inwards:

\begin{itemize}
\item At low temperatures (below 700~K), FeS (troilite) will be formed and is
stable, with the remaining iron (excess Fe over S) contained in pure iron
particles. Silicon will be in magnesium-rich amorphous silicates, while
SiO$_{\rm 2}$ is found to be unstable in chemical equilibrium.
\item At higher temperatures (around 800~K), FeS will disappear and contribute
to metallic iron, while amorphous silicates will anneal into crystalline
silicates.
\item At even higher temperatures (1300-1400~K), both crystalline silicates
and solid iron can no longer survive and are destroyed; the last remaining
dust particles are aluminium-rich species such as corundum (Al$_{\rm
2}$O$_{\rm 3}$).
\item Above 1850~K also these last remaining dust species can no longer
survive.
\end{itemize}

From a comparison with dust condensation experiments by e.g. Nuth
and Donn (1982), it is also known that also non-equilibrium
processes must play a role in the formation of dust (e.g. Tielens
et al. 2005). Unfortunately, condensation paths which also include
kinetic considerations are not yet available (e.g. Gail 2003),
mainly due to the lack of reaction rates. This is especially true
for the first step in the dust formation process in oxygen-rich
environments, the nucleation of tiny seed particles from the gas
phase.

However, the predictions based on the chemical equilibrium
considerations are already in relatively good agreement with
observations. In a more refined model, Gail (2004) added radial
mixing, moving processed material from the inner disk to more
outward regions. We should also note that several additional dust
components can exist in the temperature-density regime of
protoplanetary disks, including Al- and Ca-containing compounds
such as
hibonite (CaAl$_{12}$O$_{19}$) and spinel (MgAl$_2$O$_4$).

\subsection{Silicates}

\begin{figure}
\centering
\includegraphics[width=0.95\textwidth]{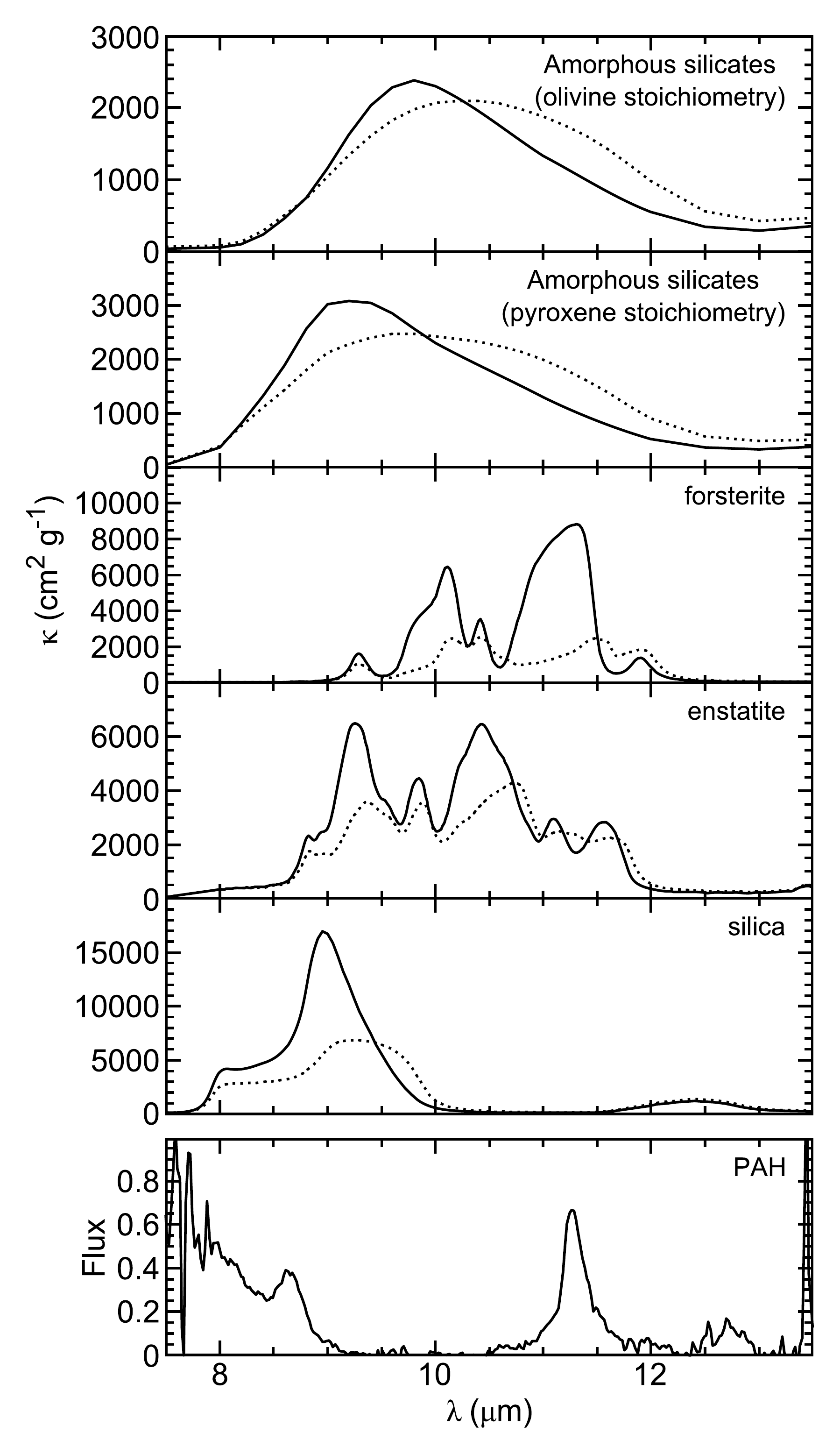}
\caption{Mass absorption coefficients of the various silicate
grains. Homogeneous spheres have been assumed for the amorphous
grains and a distribution of hollow spheres for the crystalline
forsterite, enstatite and the amorphous silica. Grains with
volume equivalent grain radii of 0.1 $\mu$m (solid lines) and
1.5~$\mu$m (dotted lines) have been used. In addition a Polycyclic
Aromatic Hydrocarbon (PAH) template is shown. After van Boekel et
al. (2005).}
\label{massabs}
\end{figure}

Silicates form a diverse class of materials, ranging from
amorphous and glass-like structures, characterized by
three-dimensional disordered networks, to well-ordered crystals
(e.g. Colangeli et al. 2003, Henning 2009). The different
silicates are assembled by linking the corners of individual
[SiO$_4$]$^{4-}$ tetrahedra through their oxygen atoms with
different levels of complexity. The negative net charge of the ion
group must be balanced by metal or hydrogen cations to produce an
electrically neutral compound. The cations are dispersed between
the individual tetrahedra or the tetrahedra arrays. Mineral
structures that have been extensively discussed in the context of
protoplanetary dust are Mg-Fe olivines and pyroxenes. Olivines
with the composition Mg$_{{\rm 2x}}$Fe$_{\rm 2(1-x)}$SiO$_{\rm 4}$
can be considered as a solid solution of their end members
forsterite (Mg$_{\rm 2}$SiO$_{\rm 4}$) and fayalite (Fe$_{\rm
2}$SiO$_{\rm 4}$). Pyroxene with the composition of Mg$_{\rm
x}$Fe$_{\rm{ (1-x)}}$SiO$_{\rm 3}$ is a solid solution formed from
enstatite (MgSiO$_{\rm 3}$) and ferrosilite (FeSiO$_{\rm 3}$). The
lattice structures of olivines and pyroxenes are very different:
olivines are island silicates (nesosilicates) with isolated
tetrahedra, while pyroxenes are chain silicates (inosilicates) in
which one oxygen atom of every tetrahedron is shared with its
neighbor. We should explicitly note that the terminology of
olivines and pyroxenes refers to crystal structures and should not
be used for amorphous silicates of the same chemical composition.
The silicate particles show a wide variety of infrared features,
characteristic of their chemical composition and structure, as is
shown for the wavelength range between 8 and 13 $\mu$m in
Fig.~\ref{massabs}, where we also show an example of a template
with PAH features.

In the equilibrium calculations for the inner disk by Gail (1998),
iron is seldom found to be incorporated in silicates, so that only
the magnesium-rich end-members of the compounds are present:
forsterite (Mg$_{\rm 2}$SiO$_{\rm 4}$) for the olivines and
enstatite (MgSiO$_{\rm 3}$) for the pyroxenes. Enstatite appears
to be the most stable of those two, with forsterite only appearing
in a narrow temperature range, just below the stability limit. It
is thus predicted that in a disk, enstatite is the dominating
component of the crystalline silicates, and forsterite is only
important in the more inward regions of the disk, near the region
where it becomes too hot to survive (around 1400~K). However, Gail
(2004) cautions that the high enstatite abundance he predicts
might need to be lowered, as the forsterite-enstatite conversion
might be too slow to reach complete chemical equilibrium within
the relevant timescales involved. We will discuss the enstatite to
forsterite ratio, as determined from observations in
Sect.~\ref{s_minera}.

The Mg/Fe ratio of the amorphous silicates located in the outer
regions of protoplanetary disks is not well-constrained. In the
Pollack et al. (1994) dust model, an average value of 0.3 was
assumed for the Fe/(Fe+Mg) ratio, guided by the mass-spectroscopy
results of comet Halley dust (Jessberger et al. 1989), as well as
results obtained for anhydrous chondritic porous interplanetary
dust particles (CP IDP; Bradley et al. 1988) that were collected
in the stratosphere. Such IDPs consist mainly of GEMS (Glass with
Embedded Metal and Sulfides), whose properties are consistent with
that of interstellar amorphous silicates (Bradley et al. 1994).
However, their origin is currently under debate: Bradley et al.
(2008) argue that they are of presolar nature, as some of the GEMS
have a non-solar isotopic composition and the ISM dust is anyway
only for a small part characterized by non-solar isotopic
composition, while Min et al. (2007) argue that most of them were
formed in the solar system. From an analysis of the shape and
position of interstellar silicate features, Min et al. (2007)
concluded that interstellar silicates are predominantly Mg-rich.
However, the interplay between shape and composition in
determining the relatively broad infrared features of amorphous
silicates introduces quite some uncertainty in such an
ana\-ly\-sis (see, e.g., Chiar \& Tielens 2006 for different
conclusions).  So far, in the analysis of protoplanetary disk
spectra, mainly amorphous silicates with a mixed Fe/Mg ratio have
been used (e.g. Bouwman et al. 2001, Kessler-Silacci et al. 2006).
Therefore, a detailed examination of high-quality spectra, taking
into account different chemical compositions and size/shape
effects still needs to be performed.

Amorphous silicates with the same stoichiometry as pyroxenes and
olivines (note again that the amorphous silicates have a different
structure from pyroxenes and olivines despite the same chemical
bulk composition) typically show two broad infrared bands at about
10 and 18 $\mu$m, corresponding to Si--O stretching and O--Si--O
bending vibrations, respectively. These bands are frequently
observed in the spectra of protoplanetary disks. The large width
of the bands results from a distribution of bond lengths and
angles, typical for the amorphous structure of these solids. The
18 $\mu$m band is additionally broadened and generally weaker due
to the coupling of the bending mode to the metal-oxygen stretching
vibrations occurring in this spectral region. The exact position
of the Si--O stretching vibration depends on the level of SiO$_4$
polymerization. As an example, the band is shifted from 9 $\mu$m
for pure (sub)micron-sized SiO$_2$ grains to about 10.5 $\mu$m for
Mg$_{\rm 2.4}$SiO$_{\rm 4.4}$ (J\"ager et al. 2003a).

In contrast to amorphous silicates, crystalline pyroxenes and
olivines produce a wealth of narrow bands from the mid-infrared to
the far-infrared wavelength range due to metal-oxygen vibrations.
In crystalline pyroxenes and olivines, the majority of the
infrared peaks are shifted to longer wavelengths with increasing
iron content.  These observed shifts are caused by an increase in
bond lengths between the metal cations and the oxygen atoms when
Mg$^{2+}$ is substituted by Fe$^{2+}$. The wavenumber shift is
very closely related with the Fe content and allows a
determination of the Mg/Fe ratio from infrared spectroscopy
(J\"ager et al. 1998). However, it is difficult to derive the
Mg/Fe ratio from 10 $\mu$m observations alone, as the shift there
is rather small; fortunately it is more pronounced for bands at
longer wavelengths, enabling the determination of the ratio. Based
on laboratory data, forsterite grains have strong bands at 10.0,
11.3, 16.3, 19.8, 23.5, 27.5, 33.5 and 69.7 $\mu$m, while
enstatite grains have bands at 9.4, 9.9, 10.6, 11.1, 11.6, 18.2,
19.3 and 21.5 $\mu$m. The exact position of these features vary
with the quality of the crystals and temperature, and will also
depend on the shape distribution of the particles.

\begin{figure}
\centering
\includegraphics[width=0.95\textwidth]{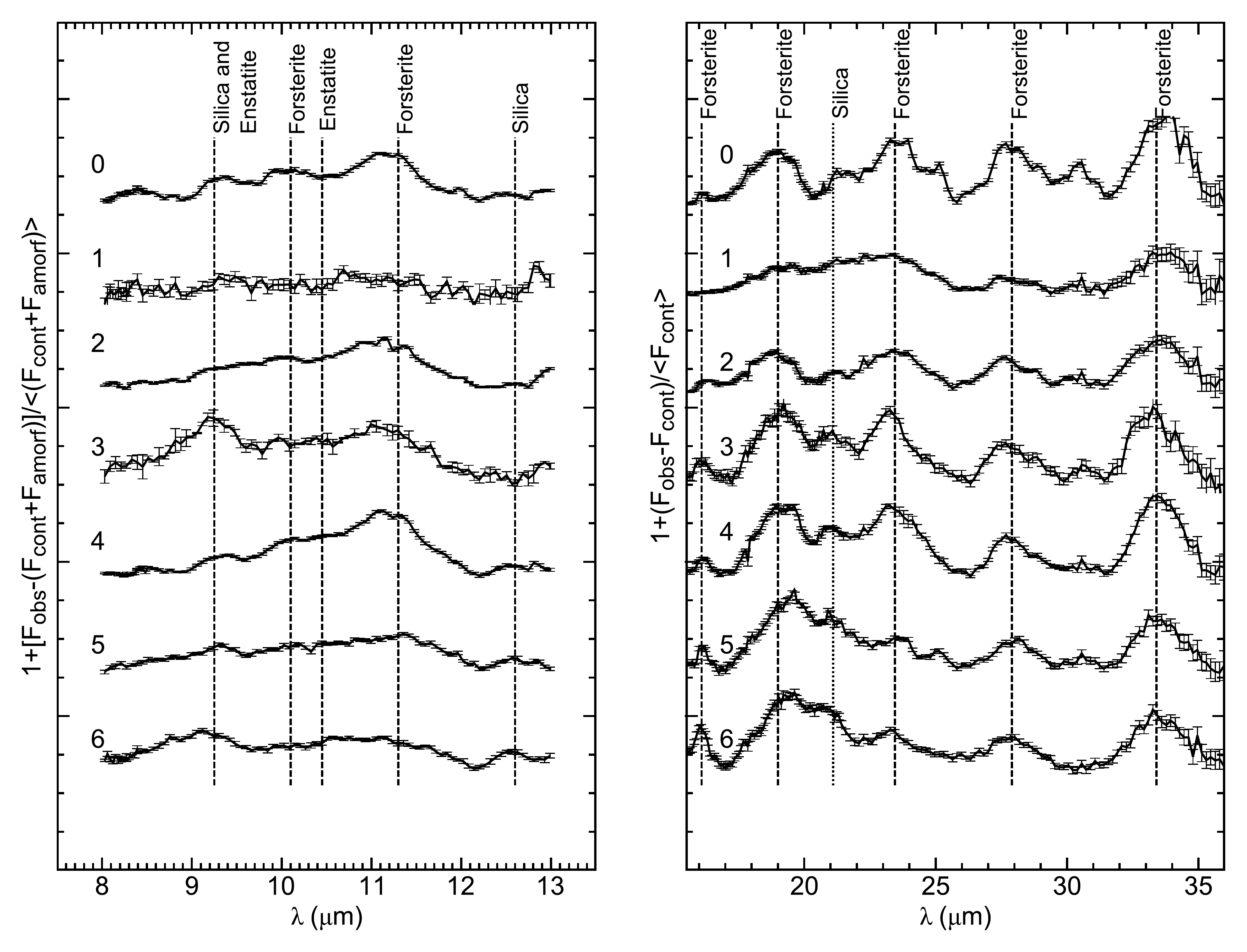}
\caption{The emission bands of crystalline silicates, as observed
with the {\it Spitzer Space Telescope} in the spectra of seven T
Tauri stars. The spectra have been normalized to a dust model fit,
thereby removing the amorphous silicate and PAH features and
enhancing the crystalline features. Note that the silica
identification is less secure. After Bouwman et al. (2008).}
\label{feps_cryst}
\end{figure}

By observing the mid-infrared wavelength range, the {\it ISO} and
{\it Spitzer} missions provided a wealth of information on the
presence of crystalline olivines and pyroxenes in protoplanetary
disks through spectroscopy (e.g. Bouwman et al. 2001, 2008;
Kessler-Silacci et al. 2006, see also Sect.~\ref{s_minera}).
Strong bands were observed at 9.3, 10.1, 11.3, 19.0, 23.4, 27.8
and 33.5~$\mu$m - in Fig.~\ref{feps_cryst}, we illustrate the
wealth of crystalline features observed in T Tauri stars.

\subsection{Carbonaceous Grains}

Carbon is a major player in the ISM, as it is a primary cooling
and heating agent. It can be present in many different forms: as
pure atoms, in simple molecules (CO, SiC, CN, CH,...) and more
complex ones (Polycyclic Aromatic Hydrocarbons; PAHs) up to
carbonaceous solids. Carbonaceous solids span a wide range of
materials (Henning \& Salama 1998, Henning et al. 2004) from
ordered structures such as graphite and diamond to complicated
amorphous structures such as a variety of hydrogenated
carbonaceous particles. The relevance of these materials as cosmic
dust analogues and their spectroscopic properties have been
summarized by Henning et al. (2004). Hydrogen-deficient amorphous
carbon grains have only weak or lacking infrared modes and will be
difficult to identify in the spectra of protoplanetary disks.
Saturated aliphatic hydrocarbons show CH stretching and
deformation modes in CH$_2$ and CH$_3$ groups at 3.4 and 6.6
$\mu$m, seen in the diffuse ISM, but so far not detected in the
spectra of protoplanetary disks. An exception is the Herbig Ae/Be
star HD 163296 where {\it ISO} spectroscopy provided evidence for
the presence of aliphatic carbonaceous dust (Bouwman et al. 2001).
We can certainly expect more evidence for this dust component from
a detailed analysis of high-quality {\it Spitzer} data (e.g.
Juh\'asz et al. 2009b).

In the Pollack et al. (1994) model for the dust composition of the
cooler, outer disk, a kerogen-like material is assumed: it is a
carbon-rich structure containing a significant amount of H, N, and
O, typical of the matrix material in carbonaceous chondrites. This
material should show C=O stretching vibrations in carbonyl groups.
However, spectroscopic evidence for the presence of this material
in disks is still lacking. So far, we have to conclude that the
nature of the carbonaceous material in protoplanetary disks
remains ill-defined.

In protoplanetary disks around Herbig Ae/Be stars, PAHs were
convincingly detected at 3.3, 6.2, 7.7, 7.9, 8.2, 8.6, 11.2, and
12.7 $\mu$m (Peeters et al. 2002, Sloan et al. 2005, Boersma et
al. 2008, Keller et al. 2008). The ratio of the band strengths can
be used to derive the charge state of the PAHs, from which the
ionization parameter can be calculated (e.g. Bakes et al. 2001).
Furthermore, the band ratio can also be related to the size of the
emitting PAHs (Allamandola et al. 1985). In T Tauri stars, there
is much less evidence for the presence of PAHs, certainly because
of the different stellar UV radiation fields (Geers et al. 2006,
2007). The presence of PAHs in the disks of Herbig Ae/Be stars
shows that at least some carbonaceous material survives in
protoplanetary disks.

Evidence for the presence of nanodiamonds have been found in a
very small number of Herbig Ae/Be stars (van Kerckhoven et al.
2002). Their infrared features at 3.43 and 3.53 $\mu$m have been
interpreted as vibrational modes of hydrogen-terminated
crystalline facets of diamond particles (Guillois et al. 1999;
pure nanodiamonds have no infrared features). Alternatively,
diamondoid molecules were recently discussed as the carriers of
these bands (Pirali et al. 2007).  There are actually only three
Herbig Ae/Be stars known to date that have clear diamond
signatures: HD 97048 (Whittet et al.  1983), MWC 297 (Terada et
al. 2001), and Elias 1 (Whittet et al.  1984). An extensive
3~$\mu$m spectroscopic survey of over 60 Herbig Ae/Be stars did
not add a single source with a pronounced diamond spectrum to the
already known objects (Acke \& van den Ancker 2006). Goto et al.
(2009) argued that the presence of diamonds in the disks around
selected Herbig Ae/Be stars may be related to the transformation
of graphitic material into diamond under the irradiation of highly
energetic particles.

Carbonaceous material was also found in solar system material: in
chondritic meteorites, nanodiamonds and graphitic particles were
detected (e.g. Sandford 1996, Hill et al. 1997), while
PAHs were observed in comets (e.g. Moreels et al. 1994,
Joblin et al. 1997).

\subsection{Iron-containing Grains}

Iron-containing particles have an important effect on the dust
opacity (Ossenkopf et al. 1992, Henning \& Stognienko 1996). In
amorphous silicate grains, they strongly influence the
near-infrared absorptivity and the temperature of the grains
(Dorschner et al. 1995). Pure iron aggregates would dramatically
increase the dust opacities (Henning \& Stognienko 1996). In the
Pollack et al. (1994) dust model, most of the iron is in silicates
and troilite (FeS), and the remainder 20\% of its solar abundance
is assumed to be in metallic iron.
In the inner disk, most of the iron would be in metallic iron. Only
at temperatures below 700 K part of the metallic iron will be
incorporated in FeS.

There is strong evidence for the widespread occurrence of Fe and
FeS in primitive solar system material, including the dust
analyzed by the mass spectrometers onboard the space probes to
comet Halley and the {\it Stardust} samples from Comet 81P/Wild 2
(e.g. Bradley et al. 1988, 1994, Zolensky et al. 2006). Solid iron
and troilite form solid solutions with the available Ni and NiS
under the conditions of protoplanetary disks and standard cosmic
element mixtures.

Despite the important role of iron in determining the opacities of
protoplanetary dust, so far no convincing spectroscopic
identification in astronomical spectra has been presented. Small
iron particles will only contribute to the general infrared
continuum and show no infrared resonance. Therefore, they cannot
be identified by distinct infrared spectroscopic features.
Laboratory measurements demonstrated that (sub)micron-sized FeS
particles should have relatively strong infrared features between
30 and 45 $\mu$m (Begemann et al. 1994, Mutschke et al. 1994).
Again, no observational evidence for the presence of these
features exists in spectra of protoplanetary disks.


\section{Optical Properties of Dust Particles}
\label{s_optprop}

In order to interpret astronomical spectra and to be able to
assign solid-state features to given species, the optical
properties of the dust particles need to be calculated. Such
calculations are based on light scattering theory and laboratory
data.

The interaction of a radiation field with a system of solid
particles can be described by their absorption and scattering
cross sections $C_{\rm abs}$ and $C_{\rm sca}$, respectively. They
describe what fraction of the incoming radiation is absorbed or
scattered by a dust particle. The extinction cross section $C_{\rm
ext}$ is given by

\begin{equation}
C_{\rm ext} = C_{\rm abs} + C_{\rm sca}
\end{equation}

The cross sections depend on the chemical and structural
properties of the solid particles, ranging from the atomic scale
(chemical composition, crystal and defect structure), to the
mesoscopic scale (porosity and inhomogeneities, mantles, surface
states), and finally the macroscopic morphology (size and shape
distribution, agglomeration, coalescence).

Instead of the extinction cross section, the mass extinction
coefficient $\kappa_{\rm m}$, defined as the extinction cross
section per unit particle mass, is often used to characterize the
extinction of light. This quantity is actually more appropriate in
describing how much light is removed from the incoming radiation
field by a fixed mass of particles. For a spherical particle,
$\kappa_{\rm m}$ = $\frac{3 Q_{\rm ext}}{4 a \delta}$ where $a$ is
the particle radius, $\delta$ the material density, and $Q_{\rm
ext}$ the extinction efficiency (extinction cross section per
geometrical cross section, $C_{\rm ext} / \pi a^2$).

The optical properties of small particles can considerably deviate
from those of bulk materials because of the occurrence of surface
modes. The structure of the interface of small particles,
including their shape, can have strong effects on their optical
behavior. A comprehensive description of the classical
electrodynamics of light absorption and scattering by small
particles goes far beyond the goal of this section and we refer to
the excellent textbook by Bohren \& Huffman (1983) for a detailed
discussion.

The qualitative features of the absorption and scattering of light
strongly depend on the ratio between the wavelength of the
incident light $\lambda$ and the size of the particle (for
a spherical particle, the radius $a$). We can distinguish three
cases:

\begin{enumerate}

\item Geometrical optics (Size parameter $x$=2$\pi$$a$/$\lambda$
$\gg$ 1): The propagation of light is described by rays which are
reflected and refracted at the surface of the scatterer and
finally transmitted, according to Snell's law and the Fresnel
formulae. The scattering of a wave incident on a particle can be
described as a combination of a reflected and a transmitted wave.
For absorbing materials, light can penetrate only within the skin
depth. Scattering, therefore, is mainly a surface effect and the
absorption cross section becomes proportional to the area of the
particle as the radius increases. In this case, the mass
absorption coefficient for a sphere scales roughly as 1/$a$. We
should note that for very large size parameters, the extinction
efficiency $Q_{\rm ext}$ approaches the limiting value 2.

\item Wave optics ($\lambda \sim a$): The angular and wavelength
dependence of the scattered radiation is dominated by
interferences and resonances. For spherical particles, this is the
domain of Gustav Mie's (1908) scattering theory, which is often
applied in astrophysics.

\item Rayleigh limit (Size parameter $x$=2$\pi$$a$/$\lambda$ $\ll$
1): If, in addition, we have $\mid m \mid x \ll 1$, where $m= n + i
k$ is the (complex) refractive index of the particle, we are in
the quasi-static limit. Then, both the incident and the internal
field can be regarded as static fields. In this regime, phase
shifts over the particle size are negligible. For non-magnetic
particles this implies that it is generally sufficient to only
consider the dipolar electric mode.

The interaction of infrared and (sub)millimeter radiation with
sub-micron sized grains can generally be considered as good
examples of the quasi-static case. However, particles with high
imaginary parts of the refractive index (metals, semiconductors,
crystalline grains) and particles of somewhat larger sizes can
easily violate the conditions for the quasi-static limit, even at
infrared wavelengths.

\end{enumerate}

\begin{figure}
\centering
\includegraphics[width=0.7\textwidth]{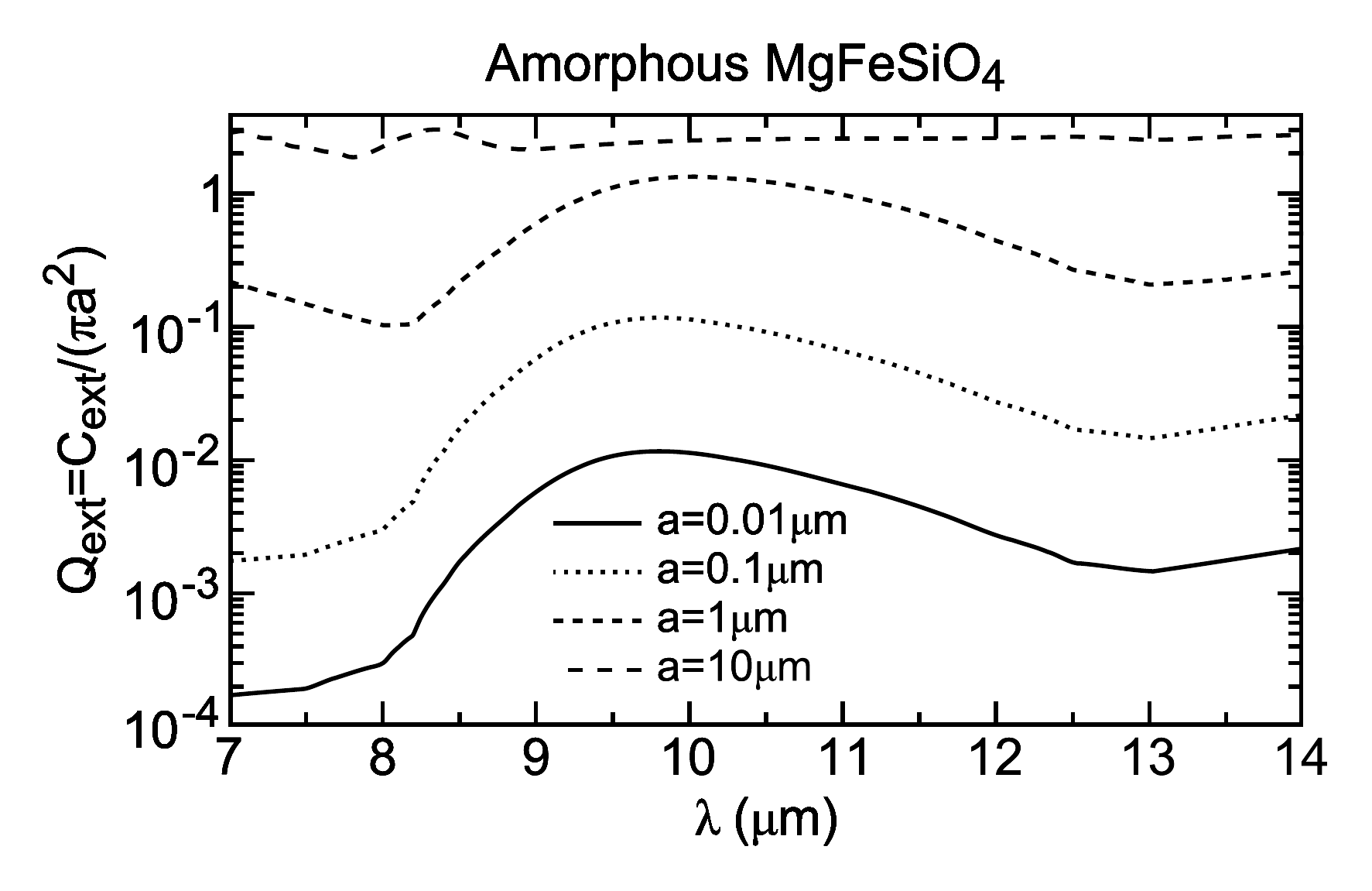}
\caption{Extinction efficiencies for spherical amorphous silicates
of different sizes. Optical constants after Dorschner et al.
(1995).} \label{fig1}
\end{figure}

In Fig.~\ref{fig1}, we show an example of the extinction
efficiencies for four different grain sizes for the case of an
infrared resonance of amorphous silicates in the 10 $\mu$m range,
caused by Si--O stretching vibrations. It is clear that the
feature changes shape with increasing grain size: larger
(micron-sized) grains show typical ``flat-topped'' features and
eventually disappears. This behavior can be used to trace the size
of the particles by infrared spectroscopy. Here, we already want
to note that the infrared features loose their diagnostic value
for grain sizes much larger than the wavelength of the feature.

For the sake of simplicity and physical insight, we will only
discuss the quasi-static case (Rayleigh limit) in the following. In
this case, there is a connection between electrostatics and
scattering by particles. Therefore, the expressions for the scattering
and absorption cross sections for (small) spherical particles can
be derived by treating the particle as an ideal dipole with the
dipole moment given by electrostatic theory (see, e.g., Bohren \&
Huffman 1998).

According to Rayleigh's law, the scattering cross section scales
with $k^4$, where $k = 2\pi/\lambda$ is the wavenumber. This means
that for very small absorbing particles (compared with the
wavelength of incident radiation) the extinction cross section is
given by the absorption cross section. For the extinction cross
section we can write

\begin{equation}
C_{\rm ext} = k {\rm Im} (\alpha)
\end{equation}

where the quantity $\alpha$ denotes the polarizability and Im
stands for the imaginary part of this quantity. The polarizability
is defined as the ratio of the induced electrical dipole moment to
the electric field that produces this dipole moment. This quantity
depends on the complex dielectric function (dielectric
permittivity), $\epsilon$ = $\epsilon_1$ + i $\epsilon_2$ (for
dielectrics $\epsilon = m^2$), of the particle and on the
dielectric function of the embedding medium, $\epsilon_m$, which
is in most cases a wavelength-dependent real number (in vacuum
$\epsilon_m$=1).

One cannot only treat spheres in the electrostatic approximation,
but also other particles, as long as their characteristic
dimensions fulfill the same conditions as defined for the
spherical particles. Therefore, we will now consider ellipsoids
as a more general particle shape, including both
spheres (all axes equal) and spheroids (two axes having the same
length).

For ellipsoids, the polarizability, $\alpha_i$, in an electric
field parallel to one of the principal axes is given by

\begin{equation}
\alpha_i = V (\epsilon - \epsilon_m)/(\epsilon_m + L_i (\epsilon -
\epsilon_m)),
\end{equation}

where $L_i$ are geometrical factors and $V$ is the volume of the
ellipsoid. The relation $L_1 + L_2 + L_3 = 1$ implies that only
two of these three factors are independent. For a continuous
distribution of ellipsoids (CDE) we get the relation (in vacuum
with $\epsilon_{\rm m}=1$)

\begin{equation}
\alpha = V (2 \epsilon/(\epsilon -1)) \log \epsilon
\end{equation}

where $\log \epsilon$ denotes the principal value of the logarithm of the
complex number $\epsilon$. The CDE in this form assumes equal probability for
the presence of every shape and averages over all orientations. This implies
that also extreme shapes are equally weighted, although they are less probable
to be present in real shape distributions. Nevertheless, the CDE can be used
for a first estimate of how important shape effects for a certain resonance
really are. It is important to note that the previous equations also
demonstrate that the mass extinction coefficient for such particles is
independent of their size, but not of their shape.

The simplest case of light scattering in the quasi-static limit is
that of a spherical particle where all axes are equal and $L_{\rm
i}$ = 1/3. This gives $\alpha_{\rm i}$ = $\alpha$ and results in
the expression

\begin{equation}
C_{\rm ext} = 4 x {\rm Im}((\epsilon - \epsilon_{\rm m})/(\epsilon
+ 2 \epsilon_{\rm m}))
\end{equation}

Equation (3) demonstrates that resonances for particles surrounded
by a non-absorbing medium occur close to the wavelength where the
imaginary part of the dielectric function is close to zero and the
real part fulfills the condition

\begin{equation}
\epsilon_1 = \epsilon_{\rm m} (1 - 1/L_i).
\end{equation}

\begin{figure}
\centering
\includegraphics[width=0.7\textwidth]{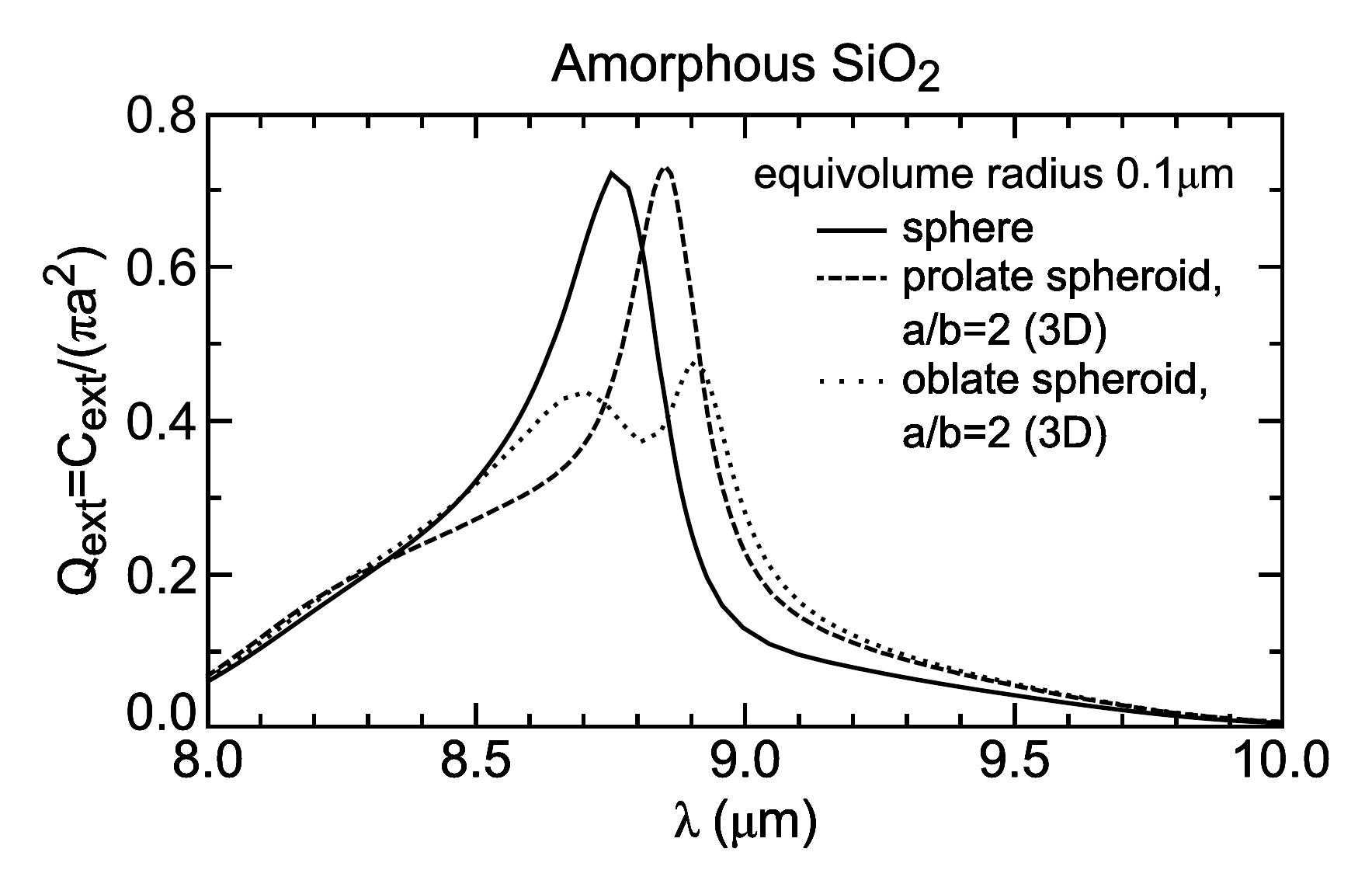}
\caption{Extinction efficiencies for silica particles  of
different shapes. Optical constants from Spitzer \& Kleinman
(1960).} \label{fig2}
\end{figure}

This equation immediately implies that the resonance wavelengths
depend on the shape of the particles. The resonances can only
occur in regions where the real part of the dielectric function is
negative (for a sphere $\epsilon_1$ = -2 $\epsilon_{\rm m}$).
Examples of astronomically relevant materials which fulfill these
criteria are SiC and SiO$_2$ in the infrared and graphite in the
UV. In contrast, the lattice features of amorphous silicates do
not always show this behavior. In Fig.~\ref{fig2} we show an
example of the shape effects for SiO$_2$ particles. In the case of
lattice modes, the resonances are always located between the
transverse and longitudinal phonon frequencies, which makes it
possible to estimate the wavelength range where the peak
absorption can occur. The equations also show that the positions
of the resonances depend on the surrounding medium. This is
important for protoplanetary dust, as it consists of core-mantle
grains where the core is made of refractory material and the
mantle is composed of molecular ices.

For particles of arbitrary shapes there exists no analytical
solution, not even in the quasi-static limit. Numerical models
frequently used for non-spherical particles are the ``Separation
of Variables method'', the ``T-matrix method'', and the ``Discrete
Dipole (or Multipole) method'' (see, e.g., Draine et al. 1988;
Michel et al. 1996; Voshchinnikov et al. 2000, 2006; Min et al.
2008).

We should stress again that the occurrence of resonances is a
property typical of small particles. For metal particles,
resonances even occur at wavelengths where the bulk material does
not show any absorption bands. In reality, the resonances will be
modified and smeared out by a distribution of shapes and the real
difficulty is to evaluate which shape distribution would be a
realistic description for the observed infrared features (Min et
al. 2005, 2007, Voshchinnikov \& Henning 2008). An analysis of the
dust composition, based on the simple assumption that the
particles are compact spheres, certainly leads to unreliable
results. Min et al. (2005, 2007) recommend the application of a
distribution of hollow spheres (DHS) for the calculations of the
dust cross sections, which is computationally easy to use. In the
Rayleigh limit, the absorption properties for a distribution of
spheroidal particles and for a distribution of hollow spheres are
very similar. However, the DHS method can also be used outside the
Rayleigh limit, in contrast to the CDE.  Absorption properties,
calculated by the DHS method, seem to provide a reasonable
representation of the shape of observed dust features (see, e.g.,
van Boekel et al. 2005, Kessler-Silacci et al. 2006).

In the dense regions of protoplanetary disks, we expect that
particles coagulate and form larger particles (Beckwith
et al. 2000, Henning et al. 2006, Natta et al. 2007). For a
description of the interaction of electromagnetic radiation with
fluffy aggregates composed of individual particles, two distinct
approaches are possible:

\begin{enumerate}

\item The ``Deterministic'' approach: The frequency-domain Maxwell
equation is solved for an individual cluster. The resulting cross
sections are calculated for many clusters and then averaged both
over the ensemble of clusters and their orientation. The advantage
of this approach is that for special systems (e.g. clusters of
spheres), exact solutions of the problem exist. For computational
reasons, however, these methods are often limited to either
comparatively small clusters or moderately absorbing systems.
Examples of this kind of approach are the discrete dipole and
multipole approximations (DDA/DMA) and the extended Mie theory for
multisphere aggregates.

\item The ``Statistical'' approach: The equations are formulated in
terms of statistically relevant quantities (e.g. average radial
density function of the clusters, density correlation function),
without any explicit treatment of individual particles. Whereas a
given cluster generally does not have any symmetry, statistical
averages show rotational invariance, unless alignment mechanisms
break this symmetry. The advantage of this approach is that only
the necessary information (ensemble and orientation averaged
quantities) enter the calculations. Examples of this approach are
the different effective medium theories and the strong
permittivity fluctuation theory.

\end{enumerate}

The dielectric function $\epsilon$ or the complex refractive index
$m$ can be determined in the laboratory for relevant astronomical
materials. The term ``optical constants'' for the real ($n$) and
imaginary part ($k$) of the complex refractive index is somewhat
misleading since the quantities strongly depend on frequency. For
certain materials they also depend on temperature. The optical
constants or dielectric functions are macroscopic quantities and
loose their meaning for small clusters and molecules.

The dispersion with frequency is determined by resonances of the
electronic system, of the ionic lattice and, at very low
frequencies, by the relaxation of permanent dipoles. In the
resonance regions the absorption becomes strong (high imaginary
part) and the real part $n$ shows ``anomalous dispersion``, i.e. a
decrease with frequency. In many cases, this behavior can be
described by Lorentzian oscillators.

Compilations of optical ``constants'' of solid materials can be
found in a number of databases which are either available in the
form of books or electronic media. The most important database in
book form is the ``Handbook of Optical Constants of Solids'',
edited by E.D. Palik, that currently consists of three volumes
which appeared in 1985, 1991 and 1998. These books are highly
recommended since they comprise detailed discussions of the origin
and errors of each data set. A database which is especially
dedicated to cosmic dust has been developed by Henning et al.
(1999) and can be found in its updated electronic form at
http://www.mpia-hd.mpg.de/HJPDOC (Heidelberg-Jena-St. Petersburg
Database of Optical Constants).


\section{Spectral Analysis Methods}

Dust properties are best studied in the infrared, as it is here
that the vibrational resonances of many astronomically relevant
materials occur. Indeed, silicates and other oxides, sulfides,
hydrogenated amorphous carbon particles, PAH molecules, and even
hydrogen-terminated nanodiamonds all show characteristic features
at those wavelengths (see Sect.~\ref{s_dust}). This makes the
infrared {\it the} fingerprint region for cosmic dust studies.
Sensitivity, spectral coverage and resolution are important
observational parameters for such studies. Ground-based
mid-infrared spectroscopy is mostly limited to the 8-13 $\mu$m
atmospheric window, but can deliver data with very high spectral
and spatial resolution; however, always with limited sensitivity
due to the thermal background of the atmosphere. The {\it ISO SWS}
spectrometer covered the extremely interesting wavelength range
between 2 and 45 $\mu$m with a spectral resolution between 1000
and 2000. In addition, the {\it ISO} {\it LWS} instrument extended
the wavelength range to wavelengths between 45 and 200 $\mu$m with
a resolution 100-200 up to 6800-9700 in high-resolution mode. The
{\it Spitzer} Short-Low and Long-Low spectrometers covered a
wavelength range between 5.3 and 14.5 $\mu$m and 14.2 and 38.0
$\mu$m, respectively, with a spectral resolution of 60-120. The
Short-High and Long-High spectrometers had a wavelength coverage
between 10.0 and 19.5 $\mu$m and 19.3 and 37.2 $\mu$m with a
spectral resolution  of 600. The {\it ISO} mission was the first
infrared space observatory which delivered high-quality data for
the bright Herbig Ae/Be stars over a wide wavelength range. The
{\it Spitzer Space Telescope}, with its unprecedented sensitivity,
provided spectra of very high quality not only for these
intermediate-mass stars but also for large and statistically
significant samples of T Tauri stars, and even made the first
measurement of spectra of brown dwarf disks possible.

\subsection{Location of the 'observable' dust grains}
\label{s_location}

The dust spectral features that are seen in emission arise in the
optically thin, warm disk atmosphere - the dust regions closer to
the mid-plane are optically thick and do not show spectral
features. This means that we can only trace a very small part of
the total disk material with infrared spectroscopy. A major
concern here is the differential sedimentation of particles:
larger grains would reach a smaller scale height, that is an
equilibrium between sedimentation and vertical turbulent mixing,
implying that through spectroscopy, we only get selective
information of the uppermost disk layer. In the case of efficient
mixing between the optically thick (featureless) disk midplane and
the upper disk layers, the observations would provide information
on the complete dust population of the disks.

The properties of the dust grains - size, shape, agglomeration
state, chemical composition and material structure - as well as
the dust temperature distribution determine the shape of the
spectral features which occur on top of the continuum, arising in
the optically thick part of the disk. In addition, the disk
geometry also has an important influence on the observed spectrum:
in disks with a flared geometry, the outer disk contributes
significantly to the spectrum in the 10~$\mu$m range, whereas in
disks with a flat geometry, the inner disk edge (the inner
``rim'') is more dominant. The inner disk edge is often strongly
``crystallized'', whereas the regions further out contain much
less crystalline material. Therefore, a disk with a flat geometry
may appear to have a higher crystallinity than a
 flared disk, even though the actual composition of the dust is identical -
this is purely a contrast effect.

Furthermore, it is important to realize that spectra around
10~$\mu$m trace different regions of the disk as a function of
(sub)stellar luminosity: $R_{10} \propto L_{*}^{0.56}$. For brown
dwarfs and T Tauri stars the emission comes from regions of 0.05
and 0.5 AU distance from the brown dwarf/star, respectively,
whereas the 10~$\mu$m emission from Herbig Ae stars traces the 10
AU range (see Kessler-Silacci et al. 2007 for a discussion of the
scaling of the silicate emission region with luminosity). In the
inner disk, a more rapid grain growth is expected, leading to
different contrast ratios between amorphous and crystalline
silicate features, as the amorphous features `flatten' with
increasing size (see Fig.~\ref{fig1}). This fact also needs to be
kept in mind when comparing spectra of objects with different
luminosities and arriving at conclusions about the amount of
crystalline grains.

\subsection{Spectral decomposition}
\label{s_decompo}

The combination of disk thermal structure and dust optical
properties can lead to degeneracies between these properties in
the modelling results, even if sophisticated radiative transfer
calculations are used to interpret the spectral energy
distributions. Such degeneracies can be reduced when additional
interferometric data, intensity maps, and polarization data are
used to further constrain the disk properties. An additional
problem comes from the fact that some materials do not show any
specific resonance (e.g. metallic iron particles in the infrared)
and that large particles ``lose'' their characteristic features.
Such ``featureless'' grains produce a continuum which is difficult
to separate from the continuum of the optically thick part of the
disk.

The most studied region in the context of dust is around 10
$\mu$m, where amorphous and crystalline silicates show features,
as well as silica and PAHs. It is fortunate that this region is
also observable from the ground, unlike most longer wavelength
regions. However, many of the features typical of crystalline
grains are located at longer wavelengths and can only be observed
by space missions. As the dust emission arises in an optically
thin region, the modelling of the features should be
straightforward, but it is complicated by the temperature
distribution of the particles and the underlying continuum. Here,
we should note that a certain wavelength range in the spectrum
corresponds to a certain temperature range in the disk. This
immediately implies that the analysis of a wider spectral range
requires the application of a temperature distribution.

In order to analyze larger data sets, especially in the 8-13
$\mu$m region, different simple spectral decomposition methods
have been used in the literature to narrow down the properties of
the grains that produce the features and to determine quantities
such as composition, shape, size, and crystallinity.

In a simple approach, the continuum below the spectral features is
modelled by a polynomial and then subtracted from the measured
spectrum. In this continuum subtraction method (Bouwman et al.
2001), the continuum is fitted outside of the feature which is
often hardly possible in ground-based observations because the
spectrum cannot be sampled in the atmospherically opaque regions.
In addition, the continuum can be associated with the bands
themselves because the grains generally contribute to the
continuum mass absorption coefficients. It is further assumed that
the dust grains have a single temperature. The
continuum-subtracted feature is then fitted with a linear
combination of mass absorption coefficients of dust particles of
different sizes, composition and structure (see Bouwman et al.
2001); the necessary optical constants for the materials are
provided by laboratory measurements.

Two other methods are very similar, but differ in the modelling of
the underlying continuum: in the 'single temperature method'  (van
Boekel et al. 2005) the spectrum is fitted by a linear combination
of optically thin and thick emission. In this approach, it is
assumed that the disk continuum is well represented by a Planck
function and that the temperature of this continuum is the same as
the temperature of the optically thin disk emission. The 'two
temperature method' (Bouwman et al. 2008) is formally identical to
the single temperature method, but fits the temperature for the
feature and the continuum separately. In the two temperature
method, the observed monochromatic flux $F_\nu$ is given by

\begin{eqnarray}
F_{\nu} &=& B_{\nu}(T_{\rm cont})C_0 + B_{\nu}(T_{\rm dust})
  \left(\sum_{i=1}^3\sum_{j=1}^5 C_{i,j} \kappa_\nu^{i,j} \right)
  + C_{\rm PAH} F_{\nu}^{\rm PAH}
\end{eqnarray}

where $B_\nu(T_{\rm cont})$ is the Planck function with a
continuum temperature $T_{\rm cont}$, $B_\nu(T_{\rm dust})$ the
Planck function with the characteristic dust temperature,
$\kappa_\nu^{i,j}$ the mass absorption coefficient for species $j$
and grain size $i$, and $F_{\nu}^{\rm PAH}$ the PAH template
spectrum. $C_0$, $C_{i,j}$ and $C_{\rm PAH}$ are weighting
factors.

The strongest limitations of these approaches are that the underlying
continuum of the optically thick dusty disk is assumed to be well represented
by a Planck function which is never the case in realistic disk models
(e.g. Men'shchikov \& Henning 1997, Chiang \& Goldreich 1997, Dullemond et
al. 2001, Dullemond \& Dominik 2004), and that even in the narrow spectral
range from 8 to 13 $\mu$m, grains of quite different temperatures contribute to
the emission.

Recently, Juh\'asz et al. (2009a) developed a more realistic but
still fast approach, the `two layer temperature distribution
method' in which a continuous distribution of temperatures is
applied, rather than a fixed temperature, and the correct
continuum is calculated. The continuum emission below the silicate
features consists of three components: a high-temperature
component from the star and from the inner rim of the disk, a
low-temperature component from the cold disk midplane and the
optically thin emission from featureless grains (e.g. carbon) from
the disk atmosphere. This again illustrates that the real
continuum is complex and cannot (and should not) be modelled by a
Planck function with a single temperature.

In the two layer temperature method, the observed monochromatic
flux $F_\nu$ is given by

\begin{eqnarray}
F_\nu = F_{\nu, {\rm atm}} &+& D_0\frac{\pi R_\star^2 B_\nu(T_\star)}{d^2} +
D1\int_{T_{\rm rim,0}}^{T_{\rm rim, min}}\frac{2\pi}{d^2} B_\nu(T) T^{\frac{2-q}{q}}dT \\
\nonumber &+& D2\int_{T_{\rm mid,0}}^{T_{\rm mid,
min}}\frac{2\pi}{d^2}B_\nu(T){T}^{\frac{2-q}{q}}dT.
\end{eqnarray}

Here $R_\star$ and $T_\star$ are the radius and temperature of the
central star, $B_\nu(T)$ is the Planck-function and $q$ is the
exponent of the temperature distribution which is assumed to be a
power law. The subscripts `atm', `rim' and `mid' refer to the disk
atmosphere, puffed-up inner rim and disk midplane, respectively.
Note that one fits the value of $q$ for the disk atmosphere, inner
rim and disk midplane \emph{separately}. $F_{\nu,{\rm atm}}$
denotes the flux of the optically thin disk atmosphere which is
given by

\begin{equation}
F_{\nu, {\rm atm}} = \sum_{i=1}^N\sum_{j=1}^MD_{i,j}\kappa_{i,j}
\int_{T_{\rm atm,0}}^{T_{\rm atm,min}}\frac{2\pi}{d^2}B_\nu(T){T}^{\frac{2-q}{q}}dT.
\label{eq:fit_td_temp}
\end{equation}

where $N$ and $M$ are the number of dust species and of grain
sizes used, respectively. The temperature range for the integrals
is fixed by the contribution from grains of different temperatures
to the emission of the actually analyzed spectral range. In other
words, the outer disk does not contribute and does not need to be
considered when analyzing the emission in the 10 $\mu$m range. In
this (and the other) method(s), one assumes that the dust mixture
is uniform over the fitted range of disk temperatures and
corresponding disk radii. This means that one has to split the
wavelength interval if one wants to analyze the whole spectral
region covered by {\it ISO} or the {\it Spitzer Space Telescope}.

Juh\'asz et al. (2009a) compared the robustness of the various
methods mentioned above with the aid of synthetic disk spectra,
calculated with a 2D radiative transfer code, so that the input
dust composition is known. They show that the two layer
temperature method does the best job in retrieving the input
composition. Furthermore, they showed that - within the interval 5
to 35 micron - the wavelength region between 7 and 17 micron is
best suited to derive the various dust parameters, assuming a dust
composition close to that generally obtained in the analysis of
disk spectra.

We note, however, that the two layer temperature method has its
own limitations, as it assumes that all dust species have the same
temperature at a given disk location and does not consider the
influence of the amount of disk flaring. This may lead to spurious
inverse correlations between the derived mass-averaged grain size
and the flaring of the disk because the radiative transfer effect,
discussed at the beginning of the section, is not appropriately
taken into account. In addition, the input grain model should
contain the major dust species expected to be present in
protoplanetary disks.

The next step in sophistication would be to use a real radiative
transfer model, but this would need additional information to
constrain the model. In the most recent models, it is now possible
to include a realistic treatment of the irradiation of the inner
disk by the central star, different flaring configurations as well
as dust sedimentation (see, e.g., Dullemond et al. 2001, Dullemond
\& Dominik 2008).


\section{Mineralogy of Protoplanetary Dust}
\label{s_minera}

The derivation of the dust properties in protoplanetary disks can
be tackled through the analysis of primitive material in the solar
system as an analogue for young disks as well as through infrared
spectroscopy of protoplanetary disks (see Henning 2003 for a
comprehensive coverage of the field of astromineralogy). In this
context, the {\it ISO} and {\it Spitzer} missions provided a
legacy of spectroscopic data on protoplanetary disks. In the
following, we will discuss what knowledge these space missions
have provided about the dust properties in disks around young
stars.

\subsection{Intermediate-mass stars: The Herbig Ae/Be stars}

Protoplanetary dust has most thoroughly been studied for disks
around intermediate-mass (2 to 8 M$_\odot$) pre-main sequence
stars, the Herbig Ae/Be stars (HAEBEs). This is mainly because
they are brighter than their lower-mass counterparts, providing
spectra with high signal-to-noise ratios. The characterization of
the dust in the disks around HAEBEs made important steps forward
with the launch of the {\it Infrared Space Observatory (ISO)},
which provided high-quality infrared spectra, and also with
ground-based data in the 8-13 $\mu$m window. Broad emission
features from silicates (at $\sim$ 9.7 and 18 $\mu$m), features
assigned to crystalline silicates, and PAH features (see Sect.
2.4) have been observed (e.g. Malfait et al. 1998, Bouwman et al.
2001, Acke \& van den Ancker 2004, van Boekel et al. 2005).

These observations showed that the dust composition and size
distribution vary widely from object to object: some objects (e.g.
AB Aur; Bouwman et al. 2000, van Boekel et al. 2005) have dust
features typical of amorphous dust grains, as found in molecular
clouds and in the diffuse interstellar medium. Other HAEBEs have
dust features showing a large fraction of crystalline dust grains,
similar to solar-system bodies such as comets and interplanetary
dust particles. The striking similarity between the silicate
mineralogy of comet Hale-Bopp's dust and the dust around the
isolated Herbig Ae/Be star HD 100546 was both an exciting and
surprising result, showing that some cometary material has seen
partial processing at high temperatures and can serve as an
analogue for silicate minerals in disks (Malfait et al. 1998).  An
amazing observation with the {\it Spitzer Space Telescope} (Lisse
et al. 2007) and ground-based N-band observations (Harker et al.
2007) was the detection of ``crystalline'' silicate features after
the Deep Impact Encounter on comet 9P/Tempel 1. That encounter
lifted grains from the cometary nucleus into the coma.  Although
the claims on the abundance of minor grain components from the
analysis of the infrared {\it Spitzer} spectrum by Lisse et al.
(2006) should be taken with some caution, the experiment clearly
demonstrates the presence of crystalline silicates in the cometary
nucleus. For a more detailed discussion on cometary grains and its
implications for dust mineralogy, heating, as well as on radial
mixing in protoplanetary disks, we refer to the papers by Hanner
(2003) and Wooden et al. (2007).

The analysis of long-wavelength spectroscopy data ($\lambda>$ 20
$\mu$m) can provide a determination of the Mg/Fe content of the
crystalline silicates as discussed in Section 2.3. Focused on the
69 $\mu$m feature, such an analysis demonstrated that crystalline
olivines in the outflows of evolved stars are predominantly
Mg-rich (Molster et al. 2002). A similar result was obtained for
the pyroxenes, based on the 40.5 $\mu$m feature. The exact
determination of the chemical composition of crystalline silicates
in the disks around young stars is more complicated, due to a lack
of relevant spectral features observed with high S/N ratio. The
only source where the 69 $\mu$m feature has been detected through
{\it ISO} observations was the bright Herbig Be star HD 100546.
Here, the analysis also showed that the crystalline silicates are
predominately magnesium-rich (e.g. Malfait et al. 1998, Bouwman et
al. 2003). The shorter-wavelength data on crystalline silicate
bands in the spectra of disks around young stars (and comet
Hale-Bopp) also point to a low iron content in crystalline
silicates (e.g. Bouwman et al. 2008, Juh\'asz et al. 2009b).

\begin{figure}
\centering
\includegraphics[width=0.75\textwidth]{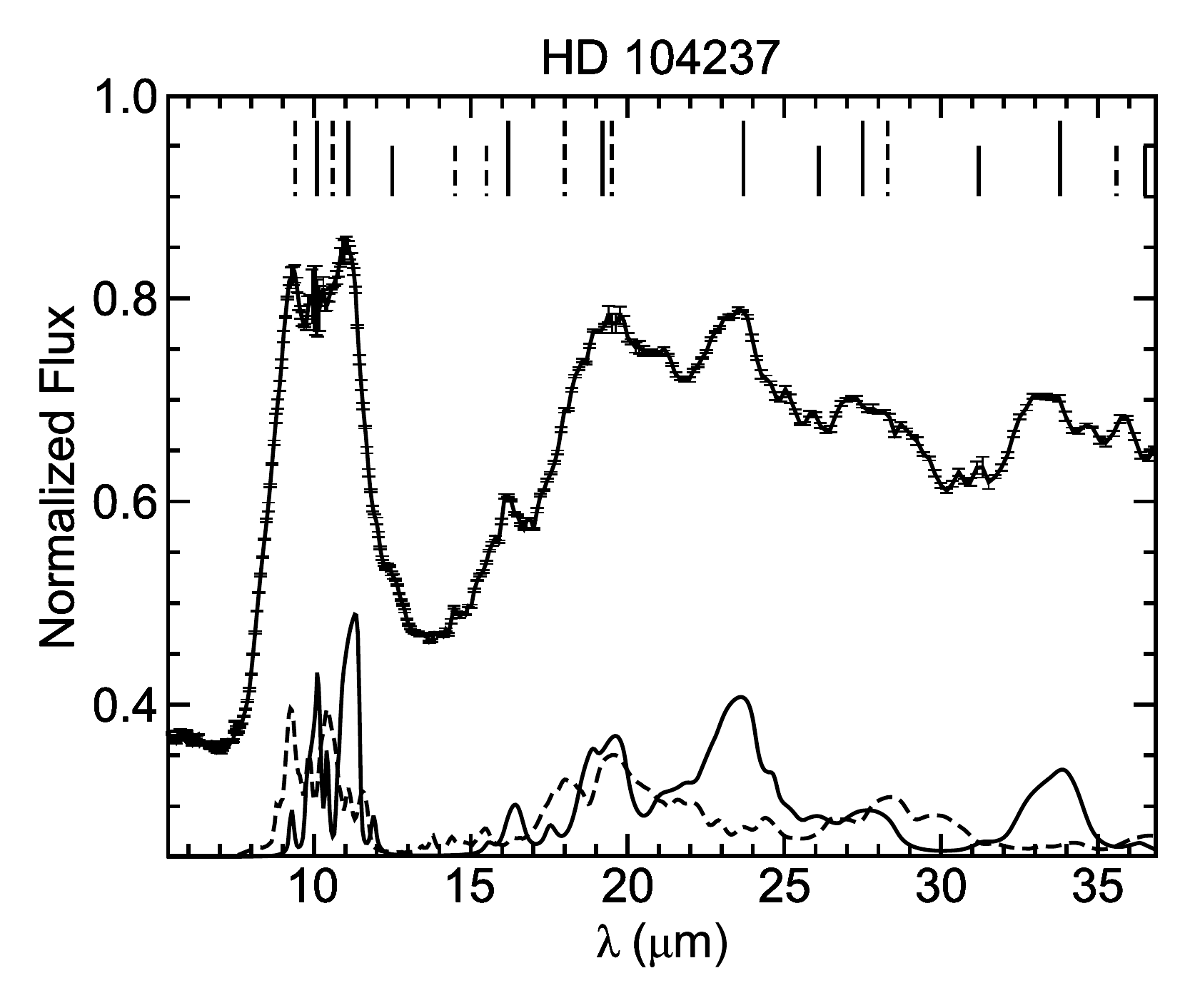}
\caption{Normalized {\it Spitzer} spectrum of the Herbig Ae star HD
104237. The positions of the forsterite (solid line) and enstatite
(dashed line) bands are indicated. In addition, the mass absorption
coefficients for 0.1~$\mu$m forsterite and enstatite grains
(distribution of hollow spheres model, volume equivalent radius) are
shown. After Juh\'asz et al. (2009b).} \label{HD104}
\end{figure}

{\it Spitzer} observations provided a compilation of high signal-to-noise
ratio spectra for about 45 HAEBE disk sources (Juh\'asz et al. 2009b). These
spectra provide a strong indication for forsterite grains dominating the dust
composition in the outer disk, characterized by the long-wavelength data.  In
Fig.~\ref{HD104}, we show the high-quality spectrum of the optically brightest
Herbig Ae star, HD~104237.

\begin{figure}
\centering
\includegraphics[width=0.65\textwidth]{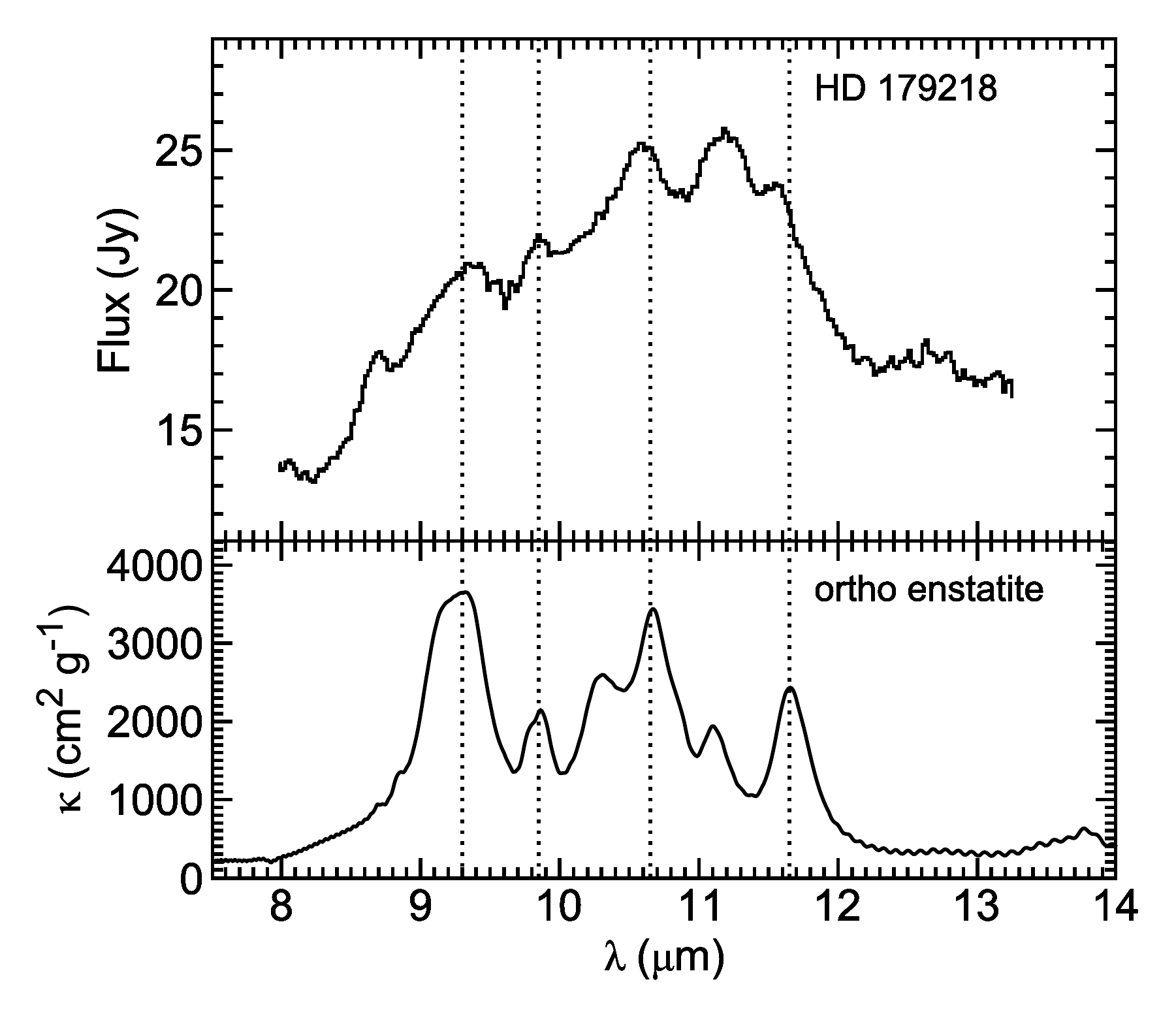}
\caption{The N-band spectrum of HD 179218 (upper panel), and the
measured mass absorption coefficients of ortho-enstatite
from Chihara et al. (2002; lower panel). The wavelengths of the
most prominent emission bands are indicated by the dotted lines.
In this object, enstatite grains are an important constituent of
the grain population that causes the 10 $\mu$m feature. After van
Boekel et al. (2005).} \label{fig4}
\end{figure}

Fig.~\ref{fig4} shows a high-quality ground-based spectrum of a
HAEBE star which illustrates that such data can
also provide important constraints on the dust composition. The
spectrum shows very convincing evidence for the presence of
enstatite in HD~179218.

Interferometric measurements with the mid-infrared instrument {\it
MIDI} at the {\it Very Large Telescope Interferometer}, providing
for the first time spatially resolved spectroscopy, demonstrate
dust processing by thermal annealing and coagulation to be most
efficient in the innermost parts of the disk ($<$ 2 AU; van Boekel
et al. 2004): a spatial gradient in amount of crystallinity and
size distribution was found in the disks of 3 Herbig Ae stars.
These data also indicated a radial dependence of the chemical
composition of the crystalline silicate dust: olivines dominate
in the inner disk, while pyroxenes dominate in the outer disk.

\subsection{The lower-mass T Tauri stars and brown dwarfs}

\begin{figure}
\centering
\includegraphics[width=0.75\textwidth]{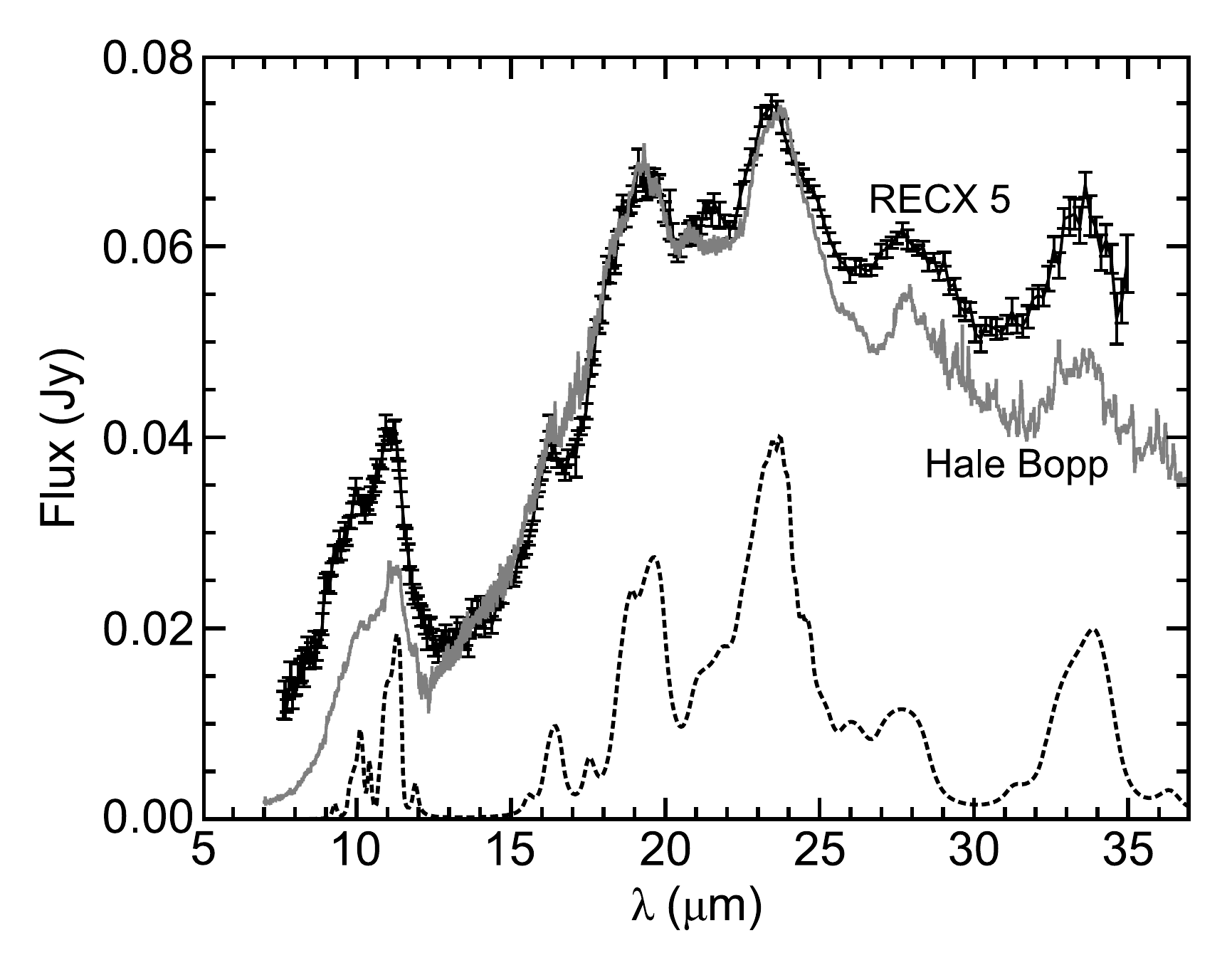}
\caption{{\it Spitzer} spectrum of the $\sim$ 9 Myr-old M4 type
star RECX5 compared with the {\it ISO} spectrum of comet Hale-Bopp
(Crovisier et al. 1997). For identification purposes, we also show
an emission spectrum (solid black line) for a distribution of
hollow forsterite spheres (volume equivalent radius of 0.1 $\mu$m)
at a temperature of 200 K.} \label{fig5}
\end{figure}

Also the grain properties in the lower-mass ($\sim$ 1 solar mass)
T Tauri stars (TTS) have been derived in the last few years, and a
similar diversity in dust properties as observed in HAEBEs was
found (e.g. Meeus et al. 2003, Przygodda et al. 2003). Thanks to
the sensitivity of {\it Spitzer}, it soon became possible to study
larger samples of TTS, confirming the similarity in dust
properties with the higher-mass stars (e.g. Forrest et al. 2004,
Kessler-Silacci et al. 2006, Sicilia-Aguilar et al. 2007, Watson
et al. 2008). Kessler-Silacci et al. (2006) also found that half
of their sample (40 TTS) show crystalline silicate features at
longer wavelengths (33-36 $\mu$m). Sicilia-Aguilar et al. (2007)
and Watson et al. (2008) found no correlation between the
crystallinity and any stellar parameter, but showed that in
general, the crystallinity in TTS is relatively low (less than
20~\%). In Fig.~\ref{fig5}, we show a comparison between the {\it
Spitzer} spectrum of the disk around a low-mass star and the {\it
ISO} spectrum of comet Hale-Bopp. The two spectra show remarkable
similarity, indicating that comet-like material is present in the
disks around TTS.

In two different samples of TTS, Bouwman et al. (2008) and Meeus
et al. (2009) determined that the forsterite to enstatite ratio is
low in the inner disk (1~AU), while forsterite dominates in the
outer (5 - 15~AU), colder regions. The same results were obtained
for the {\it Spitzer sample} of HAEBEs (Juh\'asz et al. 2009b). This
is in contradiction with the chemical equilibrium calculations by
Gail (2004), which predict that - assuming the crystalline
silicates form as high temperature gas phase condensates -
forsterite is only present in the innermost regions where
crystalline silicates can survive, while enstatite dominates in
the more distant regions. Also radial mixing, investigated in the
study by Gail (2004), would not resolve this discrepancy, as this
model still predicts enstatite to dominate the outer disk regions.
Here one should keep in mind that the reaction rates for the
conversion from forsterite to enstatite and vice versa are still
relatively uncertain.

Most likely, non-equilibrium processes contribute to the formation
of enstatite (see the extensive discussion on this topic by
Bouwman et al. 2008). As an alternative scenario for the formation
of crystals, local heating in transient events has been discussed
- either by lightning in the disks (Desch \& Cuzzi 2000), or by
shocks caused by gravitational instabilities (Harker \& Desch
2002). Compositional and structural features of enstatite and
forsterite in primitive chondrite matrices also point to the
formation through shock heating in the solar nebula, at distances
from 2-10 AU from the Sun (Scott \& Krot 2005). Here, we should
note that potential correlations between stellar parameters and
crystallinity may be erased by additional processes such as
amorphization of grains through ion irradiation associated with
stellar activity (see, e.g., J\"ager et al. 2003b).

In the mid-infrared spectra of a few T Tauri stars, taken with the
{\it Spitzer Space Telescope}, Sargent et al. (2008) detected
prominent narrow emission features at 9.0, 12.6, 20, and sometimes
16.0 $\mu$m, indicating the presence of crystalline SiO$_2$
(silica). Modeling suggests that the two polymorphs of silica,
tridymite and predominantly cristobalite, which form at high
temperatures and low pressure, are the dominant form of silica
responsible for the spectral features. This material is certainly
largely the result of processing of primitive material in the
protoplanetary disks around these stars. Tridymite and
cristobalite, once formed, must be cooled quickly enough in order
to keep their crystalline structure.

Finally, for brown dwarfs, the field of disk studies is quite new,
but also here Spitzer has somewhat lifted the veil. The silicate
emission feature observed in young brown dwarfs in Chamealeon I (2 Myr,
Apai et al. 2005) suggest similar dust processing as in the young
stellar disks: the 10 $\mu$m feature varies from source to source,
with different degrees of flattening (due to larger grains).
Remarkably - given their cooler temperature - a high degree of
crystallinity (between 10 and 50~\%) is derived.

Here, we note that the brown dwarf spectra generally have a lower
signal-to-noise ratio than those of the brighter stars, which
makes the derivation of dust parameters less reliable. In
addition, due to the low luminosity of brown dwarfs, we probe a
physically much smaller region of their disks so that we only see
those regions where grain growth is expected to be very fast (see
also Sect.~\ref{s_growth}). This may also explain the
observational evidence for often flatter disk geometries in brown
dwarfs, due to the fast settling of larger grains and the
associated transition from flared to flat disk structures. We
already noted in Sect.~\ref{s_location} that a flatter disk may
have a higher apparent crystallinity than a flared disk, even
though the actual composition of the dust is identical - this may,
at least partly, explain the brown dwarf results.

In this context, it is interesting to note that the 10~$\mu$m
feature in brown dwarfs seems to disappear fast: in the 5 Myr-old Upper
Scorpius region, the feature is either absent or very weak (Scholz et al.
2006), while the feature is completely absent in the 3 brown
dwarfs observed in TW Hya (10 Myr; Morrow et al. 2008), so that
not much is known about dust evolution in brown dwarf disks.

\subsection{PAHs and non-silicate dust}

PAH emission is widespread in disks around HAEBEs (see Sect.2.4),
with stronger features in more flaring disks, and weak or absent
features in geometrically flat disks (e.g. Acke \& van den Ancker
2004). This can be explained by the fact that flared disks
intercept a larger fraction of the UV radiation from the central
star than flat disks. However, Dullemond et al. (2007) found that
dust sedimentation can enhance the infrared features of PAHs. For
disks with low turbulence, the sedimentation causes the thermal
(larger) dust grains to sink below the photosphere, while the PAHs
still stay well-mixed in the surface layer. The sedimentation of
the larger grains would also lead to a reduced far-infrared flux.
Therefore, this investigation predicts that sources with weak
far-infrared flux have stronger PAH features, which is - at least
among the HAEBEs - opposite to what has been observed. This
suggests that sedimentation is not the only factor responsible for
the weak mid- to far-infrared excess in some disks. We also refer
to Keller et al. (2008) for a discussion of this topic.

PAH emission is also expected to be less strong in the cooler T
Tauri stars, as the PAH molecules are excited by UV photons. Geers
et al. (2007) showed that at least 8\% of TTS show PAH features,
but their latest spectral type is only G8. Bouwman et al. (2008)
report the first detection of the 8.2 $\mu$m PAH feature in young
low-mass objects, and observe this feature in 5 of their 7
sources.

It is interesting to note that apart from the discussed silicates
and PAHs, not much convincing spectroscopic evidence for other
grain components has been found - so far.

\section{Evidence for Grain Growth}
\label{s_growth}

Grain growth in protoplanetary disks is a complex process, driven
by gas-grain dynamics, leading to collisions between the particles
and finally coagulation (Beckwith et al. 2000, Henning et al.
2006, Dominik et al. 2007, Natta et al. 2007). A state-of-the art
grain growth model for the conditions of protoplanetary disks has
recently been developed, taking into account radial drift and
vertical sedimentation, as well as coagulation and the relevant
microphysics (Brauer et al. 2008).

\begin{figure}
\centering
\includegraphics[width=0.9\textwidth]{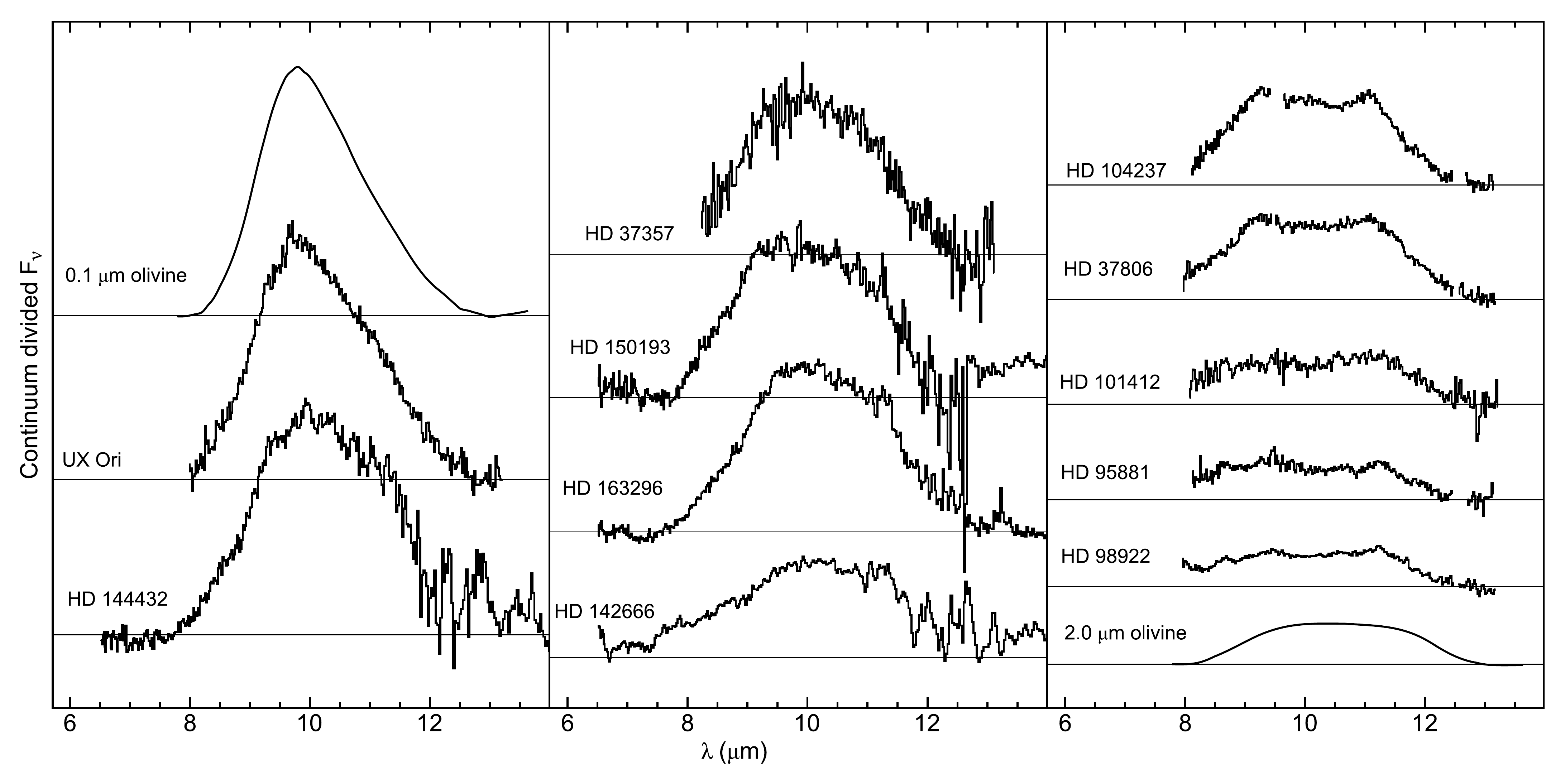}
\caption{Infrared spectra of Herbig Ae/Be stars ordered by peak
strength, illustrating the size effect on the shape and strength of
the silicate feature. After van Boekel et al. (2003).}
\label{boekel_growth}
\end{figure}

In the ISM, the mass-averaged grain size for silicates is smaller
than 0.1 micron (Kemper et al. 2004). In young disks, the derived
average grain size varies a lot between the objects (e.g. Bouwman
et al. 2001, van Boekel et al. 2005), but it is generally found to
be much larger than in the diffuse interstellar medium. Van Boekel
et al. (2003) related the shape and the strength of the 10 $\mu$m
feature, and showed that this relation provides proof for grain
growth (see Fig.~\ref{boekel_growth}): strong and triangular
10~$\mu$m features are typical of submicron-sized grains, whereas
a weaker and flattened structure indicates the presence of grains
with sizes between 2 and 4~$\mu$m. Here, we should again note that
the shape of the feature can be influenced by other parameters
than size (Min et al. 2006, Voshchinnikov et al. 2006,
Voshchinnikov \& Henning 2008). The actual particle size can be
underestimated when using homogeneous spheres instead of fractal
dust aggregates, as pointed out by Min et al. (2006).

{\it Spitzer} observations of 40 TTS by Kessler-Silacci et al.
(2006) reveal evidence for fast grain growth to micron-sized
grains in the disk surface. They did not find a correlation with
either age of the systems, nor with accretion rates.
Kessler-Silacci et al. (2006) further found that later type (M)
stars show flatter 10 $\mu$m features (pointing to larger grain
sizes) than earlier type (A/B) stars. This finding was confirmed
by Sicilia-Aguilar et al. (2007), who reported that strong
silicate features are a factor of $\sim$ 4 less frequent among
disks around M-type stars compared to stars of earlier type. This
inverse relation between stellar luminosity and grain size was
further investigated in a study by Kessler-Silacci et al. (2007).
It can easily be explained by a different location being probed at
10 $\mu$m: for later type stars, this is more inwards than for
earlier types (e.g. TTS 0.1-1 AU vs. HAEBEs 0.5-50 AU). In the
inner disk we would expect a faster grain growth because of the
higher collision rates of grains. This may also explain the fast
disappearance of the silicate features in the spectra of brown
dwarfs, discussed in the previous section.

Sicilia-Aguilar et al. (2007) found a somewhat counter-intuitive
relation for TTS in the cluster Tr37. Only the youngest (0-2 Myr)
objects show evidence for large grains by flat and weak spectral
features, while features typical of submicron-sized grains were
seen in the oldest objects ($>$ 6 Myr), indicating that dust
settling removes larger grains from the disk atmosphere. In
addition, objects with a lot of turbulence (witnessed by larger
accretion rates) have larger dust: it is likely that the
turbulence supports large grains against settling towards the disk
midplane.  They also relate the observed weakness of silicate
emission features in later-type objects with inner disk evolution.

Dullemond \& Dominik (2008) investigated the effect of
differential settling of grains on the appearance of the silicate
feature. They confirmed that sedimentation can turn a ``flat''
feature into a ``triangular'' one, but only to a limited degree
and for a limited range of grain sizes. Only in the case of a
bimodal size distribution, i.e. a very small grain population and
a bigger grain population, the effect is strong. If sedimentation
were the sole cause of the feature variation, one would expect
disks with weak mid- to far-infrared excess to have a stronger 10
$\mu$m silicate feature than disks with a strong excess at these
wavelengths, this is not what has been observed in a sample of 46
HAEBEs with a wide range of mid- to far-infrared excesses (Acke \&
van den Ancker 2004).

In a sample of 24 HAEBEs, van Boekel et al. (2005) derived the
composition and grain size of the dust with the ``single
temperature method'' (see Sect.~\ref{s_decompo}). They used this
decomposition to derive the mass fraction of the observable grains
in the 10~$\mu$m region, and found that a high crystallinity
(above 10~\%) is only observed in those cases where the mass
fraction in large grains is higher than 85~\%, suggesting that
crystallization and grain growth are related. However, this may
partly be a contrast effect, as the strength of the amorphous
feature becomes weaker with increasing grain size, thus more
clearly revealing crystalline features.

For grains with sizes larger than about 5~$\mu$m, 10~$\mu$m
spectroscopy will no longer be able to trace such particles which
anyway may sediment below the optically thin atmosphere. Here
millimeter observations of HAEBEs and TTS have provided evidence
for the presence of even cm-sized grains in protoplanetary  disks
(see, e.g., Natta et al. 2007 for a review). In the case that the
disks are optically thin at millimeter wavelengths - a feature
which can be expected because of the low mass absorption
coefficients at these wavelengths - the slope of the spectral
energy distribution ($F_{\nu}$ $\propto$ $\nu^{\alpha}$) can be
directly related to the frequency dependence of the mass
absorption coefficient ($\kappa(\nu)$ $\propto$ $\nu^{\beta}$) via
the relation $\beta$ = $\alpha - 2$. Here the Planck function is
represented by its Rayleigh-Jeans approximation. For typical
submicron-sized dust grains in the diffuse ISM, the spectral index
$\beta$ has been found to be close to 2 (e.g. Draine \& Lee 1984).
For particles that are very large compared with the wavelength,
which will block the radiation by virtue of their geometrical
cross sections, the mass absorption coefficient becomes
independent of frequency (grey behavior). Particles of about the
same size as the wavelength, for instance 7~mm, selected for
studies at the {\it Very Large Array - VLA}, i.e. pebble-sized
particles, are expected to have spectral indices in the
intermediate range. For silicate particles, $\beta$ indices $\leq$
1 suggest the presence of dust particles with millimeter sizes.
Such values have been found in the {\it VLA} studies of disks
around HAEBEs (Natta et al. 2004) and TTS (Rodmann et al. 2006),
clearly indicating the presence of large particles.


\section{Conclusions and Future Directions}

The last decade has seen huge progress in our knowledge of dust
properties in protoplanetary disks, which was made possible mostly
through observations with both the {\it Infrared Space
Observatory} and the {\it Spitzer Space Telescope}. A first
surprise was the vast variety in appearance of dust features
around young stars. A more detailed analysis, however, showed that
those dust features, although so different in appearance, could be
related to a handful of dust species, with different sizes and
structures.

The dust in protoplanetary disks bears the signature of processing
compared to ISM dust, the grains are clearly larger and most young
sources also show evidence for crystalline silicates.

Despite the large number of spectra now available to the
community, it has proven difficult to pin down the relationships
between the observed dust properties, on the one hand, and the
stellar and disk properties on the other hand. The obviously
expected relation of increasing grain size with age, or of
increasing crystallinity with higher temperature of the central
star has proven to be incorrect.

The problem lies in the fact that there are many parameters to
take into account: the age, stellar luminosity, disk flaring angle
and dust sedimentation, presence of a close companion, to name
just a few. Furthermore, dust features cannot be directly related
to a specific dust species and size: other parameters, such as the
dust temperature or the shape of the dust particles also play an
important role. In addition, good laboratory data for astronomical
dust species over the full observable wavelength range remain a
much needed ingredient. The same is true for reaction rates which
determine the conversion between different grain species.

Therefore, we expect to make more progress in the coming years by
comparing the wealth of data provided by {\it Spitzer} in a
thorough statistical study, eliminating as much as possible those
parameters that can be determined, so that, e.g., when studying
the size distribution, stars with a similar luminosity are being
compared.

In the coming years, several new observatories and instruments
will become available. With the launch of the {\it Herschel Space
Observatory} in 2009, covering the far IR wavelength region (PACS
instrument 57-210~$\mu$m) we will be able to  study the lattice
vibrations of heavy ions or ion groups with low bond energies with
a high signal-to-noise. In particular, studies of the forsterite
band at 69~$\mu$m will be promising in the context of determining
the dust temperature and the composition of the olivines, while
aqueous alteration can be studied through features of hydrous
silicates at 100-110~$\mu$m.

Further on the horizon lies the launch of the {\it James Webb
Space Telescope (JWST)}, where the unprecedented sensitivity will
allow us to study the disks of brown dwarfs with the same ease as
we now study T Tauri stars, opening up yet another region of the
parameter space in dust processing studies.

Last but not least, a lot can be learned from comparing
observations of protoplanetary disks and dust with the composition
of primitive material in the solar system, provided by the
analysis of meteoritic material and interplanetary dust particles
of cometary origin as collected by the {\it STARDUST} mission.

%
%
%



%
%
%

\printindex
\end{document}